\documentclass[journal]{IEEEtran}
\usepackage{setspace}%使用间距宏包
\usepackage{diagbox}
\usepackage{cite}
\usepackage[square, comma, sort&compress, numbers]{natbib}
\usepackage{fancyhdr}
\usepackage{color}
\usepackage{graphicx}
\usepackage{multirow}
\usepackage{color}
\usepackage{placeins}
\usepackage{float}
\usepackage{tabularx,colortbl}
\usepackage{amsmath}
\usepackage{amsfonts}
\usepackage{amssymb}
\usepackage{amsthm}
\usepackage{epstopdf}
\usepackage{cite}
\usepackage{cite}
\usepackage{amsmath,amssymb,amsfonts}
\usepackage{bm}
\usepackage{algorithmic}
\usepackage{graphicx}
\usepackage{textcomp}
\usepackage{xcolor}
\usepackage{epstopdf}
\usepackage{bm}%%黑斜体
\begin{document}
\title{MIMO Channel Information Feedback Using Deep Recurrent Network}
\author{\IEEEauthorblockN{Chao~Lu\IEEEauthorrefmark{0}, Wei~Xu\IEEEauthorrefmark{0}, Hong~Shen\IEEEauthorrefmark{0}, Jun~Zhu\IEEEauthorrefmark{0}, and Kezhi~Wang\IEEEauthorrefmark{0}}
\vspace{-1cm}
\thanks{Manuscript received October 29, 2018; accepted November 15, 2018. This work of was supported by NSFC under grants 61871109, 61871108 and 61601115, the Six Talent Peaks project in Jiangsu Province under GDZB-005, the Natural Science Foundation of Jiangsu Province under BK20150635, and the Royal Academy of Engineering under the Distinguished Visiting Fellowship scheme. The editor coordinating the review of this paper and approving it for publication was Dr. G. A. Baduge.\emph{(Corresponding authors: Wei Xu, Hong Shen.)}

C. Lu, W. Xu and H. Shen are with the National Mobile Communications Research Laboratory, Southeast University, Nanjing, China (\{220170709, wxu, shhseu\}@seu.edu.cn).

J. Zhu is with the Qualcomm Incorporated, San Diego, CA, USA (junzhu@qti.qualcomm.com).

K. Wang is with the Department of Computer and Information Sciences, Northumbria University, Newcastle, UK (kezhi.wang@northumbria.ac.uk).
}
}



%Telephone: (800) 555--1212, Fax: (888) 555--1212}


\maketitle

\newtheorem{mylemma}{Lemma}
\newtheorem{mytheorem}{Theorem}
\newtheorem{mypro}{Proposition}
\begin{abstract}
%\boldmath
In a multiple-input multiple-output (MIMO) system, the availability of channel state information (CSI) at the transmitter is essential for performance improvement. Recent convolutional neural network (NN) based techniques show competitive ability in realizing CSI compression and feedback. By introducing a new NN architecture, we enhance the accuracy of quantized CSI feedback in MIMO communications. The proposed NN architecture invokes a module named long short-term memory (LSTM) which admits the NN to benefit from exploiting temporal and frequency correlations of wireless channels. Compromising performance with complexity, we further modify the NN architecture with a significantly reduced number of parameters to be trained. Finally, experiments show that the proposed NN architectures achieve better performance in terms of both CSI compression and recovery accuracy.

\end{abstract}

% keywords
\begin{IEEEkeywords}
Channel state information (CSI) feedback, recurrent neural network (RNN), multiple-input multiple-output (MIMO).
\vspace{-0.5cm}
\end{IEEEkeywords}

% \fi
\IEEEpeerreviewmaketitle
\section{Introduction}
As a key technology in 5G, massive multiple-input multiple-output (MIMO) has become a focus in both academia and industry in recent years. Massive MIMO has shown superior performance in terms of system capacity, energy efficiency, and anti-interference capability. In order to reap the performance gain, it is critical to obtain channel state information (CSI) accurately at the transmitter.

In massive MIMO communications, the number of antennas at the base station (BS) is usually large, and thus the overhead of downlink pilots and uplink CSI feedback can be quite high. The conventional codebook based method quantizes the CSI into a number of bits \cite{AoD}. However, it can fail to achieve satisfactory performance since it can not linearly increase the performance with the quantization bits owing to its exponentially increasing complexity.
%Compressed sensing (CS) based channel estimation exploits the sparsity of some wireless channels \cite{sparse1,sparse2}.
%Compared to traditional methods, the compressed sensing (CS) based channel estimation methods can effectively reduce the pilot overhead with channel sparsity assumption. \cite{overhead1}. When the CS methods are extended to compress CSI for feedback, they can fail to achieve satisfactory performance in many cases \cite{CsiNet}.

Recently, neural network (NN) has shown great potentials in addressing some wireless communication challenges which are naturally nonlinear problems \cite{SCMA, MIMO1}.
%The authors in \cite{MIMO1} treated MIMO communication system as an encoder-decoder NN. This design, instead of exploiting customized algorithms in conventional communication systems, trained the network with data and learned network parameters in an end-to-end manner.
Deep learning based methods have achieved remarkable progress in the field of MIMO channel quantization and feedback.
By treating MIMO channel matrix as an image, the network, namely CsiNet in \cite{CsiNet}, adopted trendy image processing network architectures, e.g., ResNet \cite{Resnet}, for MIMO CSI compression and feedback.

%In many applications of wireless communication, MIMO channel generally presents correlations across time and frequency domains. %These correlations are helpful in enhancing potential performance of CSI compression in a time-varying channel.
%In this letter, we propose to introduce compression/uncompression modules with memory into deep network, which has the ability to exploit channel correlations and achieve notable performance enhancement.
In this letter, we propose a new NN
%structure to improve the existing method in \cite{CsiNet}
by incorporating recurrent neural network (RNN) to catch the temporal channel correlation.
Compared to the parallel work in \cite{CsiNetLSTM}, which also exploits RNN to enhance the network in \cite{CsiNet}, our work focuses on the design of feature compression and uncompression modules while \cite{CsiNetLSTM} considers enhancing the channel recovery module.  The compression and uncompression modules in \cite{CsiNet} and \cite{CsiNetLSTM} both utilize linear fully-connected networks (FCN), which are not sufficiently effective especially in tracking the temporal correlations in compression.
%The conventional method in \cite{AoD} (re)stores the CSI as an index word while our methods compress the channel into a vector directly.

The main contributions of this work are summarized as follows:

\begin{itemize}
\item We develop a deep NN using modules with memory for CSI compression and feedback in MIMO communications. Recurrent compression and uncompression modules are designed to exploit the channel correlation effectively. The performance is significantly improved compared to existing methods.
\item We further devise a method to reduce the training complexity. The number of parameters in the network reduces sharply while the performance advantage still retains.
\end{itemize}
%The rest of the paper is organized as follows.
%In Section II, we describe the mmWave MIMO channel model and the framework of deep learning based channel feedback.
%Section III presents the proposed recurrent NN architecture and further modifies the proposed architecture with a reduced parameter size.
%Section IV evaluates the proposed architecture with numerical comparisons.

\section{CSI Compression and Feedback}
%\subsection{System Model}

\begin{figure*}[thbp]
\centerline{\includegraphics[width=17cm,height=3.5cm]{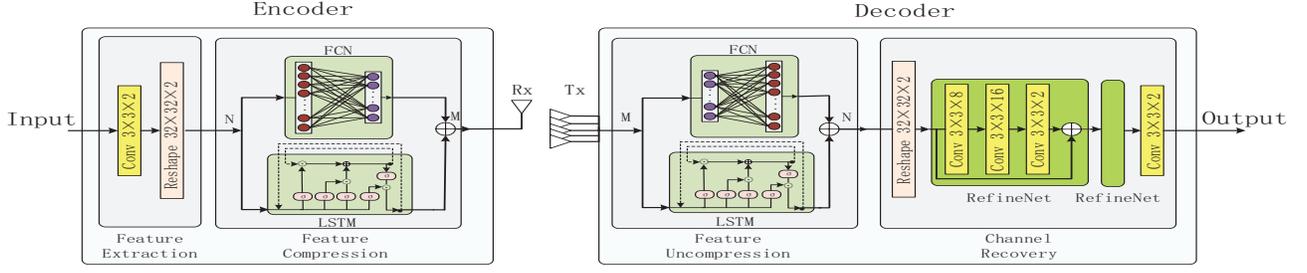}}
\caption{Architecture of encoder-decoder network, the given feature extraction module and the channel recovery module following the settings of \cite{CsiNet}.}
\label{encoder_decoder}
\end{figure*}

%We consider a frequency division duplex (FDD) mmWave MIMO downlink with $N_t$ antennas at the BS and a single antenna at each user (UE). The received signal at the $n$th subcarrier can be expressed as:
%\begin{equation}\label{eq:ReveivedSignal}
%y_n={{\tilde{\bf{h}}}^H_n}{{\bf{v}}_n}x_n+z_n,
%\end{equation}
%where ${\tilde{\bf{h}}}_n \in {\mathbb{C}}^{N_t \times 1}$ represents the channel response at the $n$th subcarrier, ${\bf{v}}_n$ denotes the precoding vector at the $n$th subcarrier, $x_n$ denotes the modulated data symbol, and $z_n$ is the additive noise.

%By applying the transformation of channel from time domain into the angular-delay domain, we have \cite{FundWireless}:
%\begin{equation}\label{eq:DFT}
%{\bf{H}}={{\bf{F}}_a}{\tilde{\bf{H}}}{\bf{F}}^H_b,
%\end{equation}
%where ${\tilde{\bf{H}}}={[{\tilde{\bf{h}}}_1, {\tilde{\bf{h}}}_2,...,{\tilde{\bf{h}}}_{{N}_c}]}^H $ is the stacked channel matrix, and $\bf{F}_a$ and $\bf{F}_b$ are the discrete Fourier transform (DFT) matrices with sizes of $N_c \times N_c$ and $N_t \times N_t$, respectively. Here $N_c$ represents the number of subcarriers.

%\subsection{CSI Compression and Feedback}
%Considering channel sparsity in MIMO, existing CS based channel estimation methods \cite{sparse1,sparse2} can be used to obtain an estimate of $\bf{H}$ in (\ref{eq:DFT}).
Consider a frequency division duplexed (FDD) MIMO downlink with $N_t$ antennas at the BS and a single antenna at each user equipment (UE). The received signal at the $n$th subcarrier can be expressed as:
\begin{equation}\label{eq:ReveivedSignal}
y_n={{\tilde{\bf{h}}}^H_n}{{\bf{v}}_n}x_n+z_n,
\end{equation}
where ${\tilde{\bf{h}}}_n \in {\mathbb{C}}^{N_t \times 1}$ represents the channel response at the $n$th subcarrier, ${\bf{v}}_n$ denotes the precoding vector at the $n$th subcarrier, $x_n$ denotes the modulated data symbol, and $z_n$ is the additive noise.
%We assumed that the estimate of ${\tilde{\bf{h}}}_n$ has been acquired.
An estimate of $\tilde{\bf{h}}_n$ is assumed to be acquired at each UE. Then, the channel is quantized with a codebook and the quantization information is sent back to the BS for CSI recovery \cite{AoD}. However, the codebook based quantization may fail to fully exploit the sparsity nature of channels since the quantization have to catch the amplitude and phase of every path in finite number of codewords. Additionally, the size of codebook grows exponentially with the number of quantization bits which can cause large storage and computational overhead. Different from the codebook based method, the NN is able to compress the CSI within short time and achieve high performance \cite{CsiNet}. Thus, in our study, we consider the application of NN, which follows the design philosophy previously proposed in \cite{CsiNet}.

In Fig. \ref{encoder_decoder}, we depict the NN based CSI compression and feedback model.
For the ease of tractability as in \cite{CsiNet}, the channel is first transferred into the angular-delay domain by 2D discrete Fourier transform (DFT) before being processed by the NN. Let ${\bf{H}}\in {\mathbb{C}}^{N_c \times N_t}$ be the channel information in the angular-delay domain, where $N_c$ denotes the number of subcarriers.
For the convenience of NN processing, $\bf{H}$ is separated into real and image parts, and all the entries are normalized within [0,1].
Instead of using codebook based CSI quantization, $\bf{H}$ is compressed through an ``Encoder" which is realized via a deep NN. It compresses the input of CSI estimate into a few bits which are then fed back to the BS.
After getting the full-bit information, a ``Decoder" at the BS recovers the CSI using another deep NN.
The output size of the Encoder and the input size of the Decoder are both set to $M$. The number of the elements of the cropped channel $\bf{H}$ is denoted by $N$. Parameter $\gamma \triangleq M/N$ represents the channel information compression ratio.

The existing Encoder design of CsiNet in \cite{CsiNet} employs a $3 \times 3$ convolutional filter and uses the leaky rectified linear unit (LeakyRELU) as the nonlinear activation function before a linear FCN to complete CSI compression.
On the other side, a linear FCN and two RefineNet modules followed by a convolutional layer constitute the Decoder. The RefineNet consists of 3 convolution operations and a jump connection. According to the basic module of ResNet \cite{Resnet}, this architecture promises the stability of the network and helps restore the channel information effectively.
%The deep NN has shown remarkable potential of improving the performance. By transforming the channel matrix to an image, image processing network can be used to deal with this problem concisely. CsiNet applied a convolutional network on a single channel matrix.

Since the channel varies slowly in many typical massive MIMO applications \cite{TimeVaryChannel1}, the correlation within a sequence of channel information can be explored for more efficient CSI compression. %In order to compress the slowly varying CSI as much as possible, we design the recurrent compress and uncompress blocks which have the storage function.
In the following section, we propose a new design of Encoder and Decoder architectures by introducing modules with memory.
Different from CsiNet, the extracted features are compressed by taking, e.g., temporal correlation, into account. Correspondingly, the proposed Decoder block at the BS also exploits the temporal correlation for CSI sequence recovery.
Mathematically, let $H_{t}(i,j) = H_{t-1}(i,j) + \alpha  |H_{t-1}(i,j)| e$, where $H_{t}(i,j)$ denotes the $(i,j)$-th entry of the $t$th channel matrix in the channel sequence, $e$ denotes the standard Gaussian random variable and $|\cdot|$ is the absolute value function. The correlation between adjacent CSI inputs is represented by $\alpha$.

\section{Proposed CSI Neural Network with Memory}
In this section, we describe the design details of our proposed NN involving compression and uncompression modules with memory.
\subsection{Recurrent Compression and Uncompression Modules}
%Following the encoder-decoder idea, we modularize Encoder and Decoder in the proposed NN. In this way, the architecture can be designed regularly owing to its modularity.
In this subsection, we modularize the Encoder and Decoder architecture in the proposed NN.

As depicted in Fig. \ref{encoder_decoder}, Encoder contains two modules.
%feature extraction module and feature compression module.
%Evidently, the
The feature extraction module mainly extracts channel features through convolution operations.
The feature compression module then compresses the features.
Correspondingly, Decoder also has two main modules.
%: feature uncompression module and recovery module.
The feature uncompression module is responsible for recovering compression features and the recovery module then restores the channel matrix.

To be more specific, the feature extraction module employs a $3 \times 3$ convolutional filter, which is referred from CsiNet \cite{CsiNet}, while the recovery module utilizes two RefineNets and a convolutional layer. The compression and uncopression modules are proposed using the long short-term memory (LSTM) network \cite{LSTM}, which has the memory function and thus can capture and extract inherent correlations, e.g. temporal correlations within input sequences. In our design, see Fig. \ref{encoder_decoder}, the input of the compression module is split into two parallel flows: an LSTM network and a linear FCN.
The FCN serves as a jump connection which can accelerate the convergence and reduce the vanishing gradient problem \cite{Resnet}. The input size of the compression module is $N$ while the output size is $M$, typically $N>M$. Since it projects the size from $N$ to $M$, we can add the two flows together at the output end. In the proposed NN architecture, we let the LSTM network learn the residual features, instead of learning correlation features directly, which is more robust.

Correspondingly at the BS, the uncompression module includes two flows which are realized by a linear FCN and an LSTM network, respectively. The only difference between the compression and uncompression modules is the input and output sizes, which needs to be symmetric.

According to \cite{LSTM}, the architecture in LSTM is shown in Fig. \ref{lstm} and the computations involved in the LSTM are:
\begin{figure}[t!]
\centerline{\includegraphics[width=5.2cm,height=2.5cm]{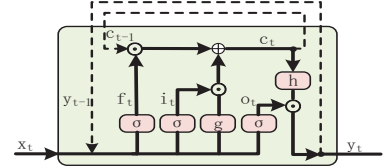}}
\caption{The LSTM cell structure.}
\label{lstm}
\end{figure}

\begin{subequations}\label{eq:LSTM}
\begin{align}
%\left[i_t,\tilde{C}_t,f_t,o_t\right] = \left[\sigma,g,\sigma,\sigma\right]([W_y,W_x]
%\left[
%\begin{matrix}
%y_{t-1}\\
%x_t
%\end{matrix}
%\right]
% + b)\\
i_t = \sigma(W_{yi} y_{t-1} + W_{xi} x_t + b_i ), \\
c_t = f_t \odot c_{t-1} + i_t \odot \tilde{C}_t, \\
y_t = o_t \odot h(c_t),
\end{align}
\end{subequations}
%\begin{subequations}\label{eq:LSTM}
%\begin{align}
%i_t = \sigma(W_{yi} y_{t-1} + W_{xi} x_t + b_i ) \\
%\tilde{C}_t = g(W_{y{\tilde{C}}} y_{t-1} + W_{x{\tilde{C}}} x_t + b_{\tilde{C}} ) \\
%f_t = \sigma(W_{yf} y_{t-1} + W_{xf} x_t + b_f) \\
%c_t = f_t \odot c_{t-1} + i_t \odot \tilde{C}_t \\
%o_t = \sigma(W_{yo} y_{t-1} + W_{xo} x_t + b_o) \\
%y_t = o_t \odot h(c_t),
%\end{align}
%\end{subequations}
where $x_t$ and $y_t$ denote the input and the output of the LSTM cell, respectively. The computations of $\tilde{C}_t, f_t$ and $o_t$ are almost the same as $i_t$ \cite{LSTM}.
$W$ and $b$ serve as the weight parameters and the corresponding bias parameters, respectively. Here $\odot$ denotes the Hadamard product. $\sigma$ and $h$ are nonlinear activation functions.
%$\sigma(\cdot)$, $g(\cdot)$ and $h(\cdot)$ are nonlinear activation functions.

The LSTM network is usually used for sequence modeling, due to its ability to capture correlation \cite{LSTM}. This can be verified through Eq. (\ref{eq:LSTM}): $c_t$ is determined by its previous state $c_{t-1}$ and the present input $i_t$. The amount of information kept and forgotten by LSTM is determined by the learned parameters. This memory mechanism enables LSTM to compress the temporal redundancy.
Regarding the training complexity, the compression and uncompression modules occupy the majority of the total parameters to be trained in our proposed NN. As for the $3\times 3\times 2$ convolutional operation, it only has 18 parameters, which is ignorable. Considering the parameters at the UE in Fig. \ref{encoder_decoder}, the FCN has $N\times M$ parameters. The parameters of the LSTM is made up of $W$ and $b$. The sizes of $W_{y}$, $W_{x}$ and $b$ are $N\times N, N\times M$ and $N$, respectively. Thus the number of parameters at the BS amounts to $(NM) + (4N^2+4NM+4N)$. Similarly, the amount of training parameters of the NN at the UE is of the same order of magnitude.
Note that, in deep learning networks, gated recurrent unit (GRU) is an alternative of LSTM for modeling sequences with memory. In our proposed NN architecture, the LSTM can be replaced by GRU without much additional changes to our current design.

\subsection{Recurrent Compression and Uncompression Modules with Reduced Complexity}
\begin{figure}[t!]
\centerline{\includegraphics[width=9.0cm,height=3.0cm]{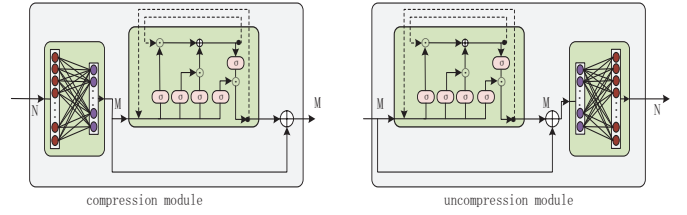}}
\caption{Recurrent compression and uncompression modules in PR-RecCsiNet.}
\label{proposed2}
\end{figure}

The above proposed recurrent CsiNet, referred to as RecCsiNet, suffers from the problem of a large number of training parameters. Here, we provide a far more effective solution to address this issue. The parameter-reduced recurrent CsiNet (PR-RecCsiNet) utilizes new compression and uncompression modules as illustrated in Fig. \ref{proposed2}. We use a linear FCN to project $M$-dimensional input to $N$-dimensional output and the output size of LSTM is reduced to $M$ in the uncompression module. Therefore, the parameter size of the uncompression module at the BS reduces to $(NM)+(4M^2+4M^2+4M)$. Another benefit of this projection is to force the LSTM to have the same input and output size. Consequently, the jump connection on the LSTM does not need any dimension transformation before the addition operation. The same modifications are correspondingly made in the compression module.

Comparing the proposed architectures in Fig. \ref{encoder_decoder} and Fig. \ref{proposed2}, RecCsiNet connects LSTM and FCN in parallel while PR-RecCsiNet connects LSTM and FCN in serial. The FCN in the two networks play different roles. FCN in RecCsiNet serves as a jump connection with dimension transformation between the input and output of LSTM. In PR-RecCsiNet, the input and output of LSTM already have the same size, which allows us to link them together directly. The FCN projection in PR-RecCsiNet is designed to reduce the training parameter size of the network by reducing the input size of LSTM.

Since the input size of the LSTM cell is reduced, the parameter size of the network is reduced notably. As we can see in Table \ref{TABLE: Parameters}, the parameter size of RecCsiNet is about 19M while the parameter size of PR-RecCsiNet is reduced to 0.8M at compression ratio 1/16. With the involvement of the projection layer, the parameter size of our recurrent network, as compared in Table \ref{TABLE: Parameters}, remains comparable with that of the existing techniques, e.g., CsiNet and CsiNet-LSTM. Since FCN compresses the features extracted from the channel matrix to a lower dimension, it could bring some information loss compared to RecCsiNet.
%However, the correlation is still well explored by the LSTM module, thus PR-RecCsiNet can still retain a significant part of the performance gain over existing methods.

\begin{table}[t!]
\scriptsize
\setlength{\abovecaptionskip}{0pt}
\setlength{\belowcaptionskip}{-10pt}
  \centering
  \caption{Number of Parameters}\label{TABLE: Parameters}
  \begin{tabular}{|c|c|c|c|}
  \hline
  \diagbox{Method}{$\gamma$}  & 1/16 & 1/32 & 1/64 \\
  \hline
  CsiNet \cite{CsiNet} & 530,656 & 268,448  & 137,344 \\
  \hline
  CsiNet-LSTM \cite{CsiNetLSTM} & 102,009,892 & 101,354,404 & 101,026,660 \\
  \hline
  RecCsiNet & 19,478,584 & 18,118,392 & 17,450,584 \\
  \hline
  PR-RecCsiNet & 793,144 & 333,816  & 153,304 \\
  \hline
  \end{tabular}
\end{table}

\section{Numerical Results}\label{sec:simulation}
In this section, we describe the detailed setup of our experiments and compare our method with existing methods.
The values of $\bf{H}$ are generated based on the COST2100 \cite{COST2100} and mmWave \cite{SCM, mmWave} channel models.
The sequence length $T$ is set to 4 for convenience.
The BS uses $N_t=32$ antennas and $N_c=1024$ subcarriers. Since the multipath arrivals are limited to short adjacent time slots around the direct path, only the first ${\tilde{N}}_c$ rows of $\bf{H}$ are non-zeros. Thus we reserve the first ${\tilde{N}}_c=32$ rows of the channel $\bf{H}$.
%, ${\tilde{N}}_c$ is set to 32.

All the networks are trained end-to-end by using the criterion of minimizing the mean squared error (MSE). The loss function is expressed as:
\begin{equation}\label{eq:MSE}
\begin{split}
L(\theta_e; \theta_d) = &\frac{1}{KT} \sum_{n=k}^{k+K-1} \sum_{t=1}^{T} \sum_{i=1}^{{\tilde{N}}_c} \sum_{j=1}^{N_t} \\
                        &| f_d(f_e(H_{n,t}(i,j); \theta_e); \theta_d) - {H}_{n,t}(i,j))|^2,
\end{split}
\end{equation}
where $\theta_e$ and $\theta_d$ denote the parameters of Encoder and Decoder. $f_e(\cdot)$ and $f_d(\cdot)$ denote the network functions of Encoder and Decoder. $H_{n,t}$ is the $t$th step of the $n$th sample, $k$ is the start index of the mini-batch, $K$ denotes the batch size, and $T$ is the sequence length.
%In this experiment, the learning rate is set to 0.01, $K=40$ and $T=4$.
The training, validation and testing sets have 100,000, 30,000 and 20,000 sequences, respectively.
The correlation factor $\alpha$ is identical in the training, validation and testing sets.
We keep $M_1=M_2$ in CsiNet-LSTM \cite{CsiNetLSTM} for a fair comparison, which implies that CsiNet-LSTM feeds back the same size of codewords in every step.
Results for both unquantized and quantized codewords are given. For the quantized version, we add a ``\emph{tanh}'' function to map all the elements to (-1,1) before the codewords are fed back, and then divide the interval uniformly by 8 bits. Thus the codeword can be replaced by a 8-bit sequence. To compare with the conventional AoD codebook based method \cite{AoD}, we test two cases with 8 bits and 16 bits for each main path. The compression ratios of the networks are accordingly set to 1/64 and 1/32.
%We keep the proposed NN be trained upto 500 epoches on the training dataset, and select final NN parameters based on the performance on the validation set.

\begin{figure}[t!]
\centerline{\includegraphics[width=5.6cm,height=2.7cm]{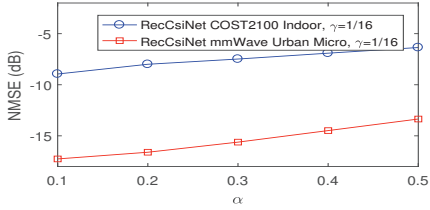}}
\caption{NMSE with different correlation coefficients without quantization.}
\label{correlation_coe}
\end{figure}

\begin{figure}[t!]
\centerline{\includegraphics[width=5.6cm,height=2.7cm]{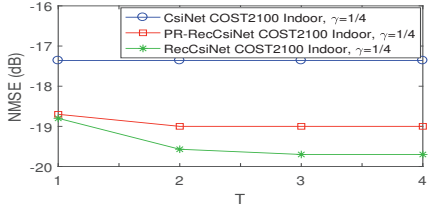}}
\caption{NMSE at different time steps without quantization and $\alpha=0.1$.}
\label{time_step}
\end{figure}

NMSE is used as the performance metric which is defined as ${\text{NMSE}} = {\mathbb{E}} \left\{ {||{\bf{H}} - \hat{\bf{H}}||^2} / {||{\bf{H}}||^2} \right\}$,
where ${\mathbb{E}}\left\{\cdot\right\}$ denotes the expectation operation and $||\cdot||$ denotes the Frobenius norm. Fig. \ref{correlation_coe} shows the performance of RecCsiNet at compression ratio of 1/16. When the correlation coefficient $\alpha$ increases, the channel temporal correlation decreases, and thus the NMSE grows correspondingly. Fig. \ref{time_step} illustrates the NMSE performance at different time steps. It can be seen that the NMSE of our networks drops noticeably before convergence as the temporal correlation appears.

\iffalse
\begin{table}[t!]
\scriptsize
\setlength{\abovecaptionskip}{0pt}
\setlength{\belowcaptionskip}{-10pt}
  \centering
  \caption {NMSE performance without quantization (\upshape{d}B), $\alpha=0.1$}\label{TABLE:nmse}
  \begin{tabular}{|c|c|c|c|c|c|}
  \hline
   & {$\gamma$} & 1/4 & 1/16 & 1/32 & 1/64 \\
  \hline
  \multirow{4}*{COST2100 Indoor}
                        & CsiNet \cite{CsiNet} & -17.36 & -8.65 & -6.24  & -5.84 \\
                        \cline{2-6}
                        & CsiNet-LSTM \cite{CsiNetLSTM} & -11.82 & -6.09 & -4.67 & -2.46 \\
                        \cline{2-6}
                        & RecCsiNet & \bf{-19.43} & \bf{-10.47} & \bf{-8.95} & \bf{-7.01} \\
                        \cline{2-6}
                        & PR-RecCsiNet & \bf{-18.97}  & \bf{-8.91} & \bf{-8.29}  & \bf{-5.97} \\
  \hline
  \multirow{4}*{mmWave Surban Micro}
                         & CsiNet \cite{CsiNet} & -8.75 & -4.51 & -2.81  & -1.93 \\
                         \cline{2-6}
                         & CsiNet-LSTM \cite{CsiNetLSTM} & -6.69 & -2.51 & -0.52 & -0.22 \\
                         \cline{2-6}
                         & RecCsiNet & \bf{-13.13} & \bf{-7.28} & \bf{-5.43} & \bf{-4.35} \\
                         \cline{2-6}
                         & PR-RecCsiNet & \bf{-10.26} & \bf{-5.32} & \bf{-3.19} & \bf{-1.98} \\
  \hline
  \end{tabular}
\end{table}
\fi

%(\upshape{d}B)
\begin{figure}[t!]
\centerline{\includegraphics[width=5.6cm,height=2.7cm]{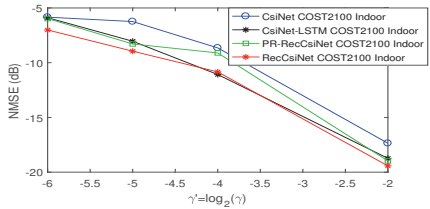}}
\caption{NMSE without quantization and $\alpha=0.1$.}
\label{cost}
\end{figure}

%(\upshape{d}B)
\begin{table}[t!]
    \scriptsize
    \setlength{\abovecaptionskip}{0pt}
    \setlength{\belowcaptionskip}{-10pt}
      \centering
      \caption {NMSE (\upshape{d}B) with Quantization MmWave Urban Micro, $\alpha=0.1$}\label{TABLE:nmse2}
      \begin{tabular}{|c|c|c|c|c|c|}
            \hline
                              & AoD codebook\cite{AoD} & CsiNet \cite{CsiNet}  & PR-RecCsiNet  & RecCsiNet\\   %& PR-RecCsiNet \\
            \hline
            {8 bits/path}     & -7.69           & -10.55                 &  \bf{-11.54}                       & \bf{-12.02}\\   %& \bf{-8.1}    \\
            \hline
            {16 bits/path}    & -9.56           & -13.99                &  \bf{-16.34}                        & \bf{-17.21}\\  %& \bf{-13.1}   \\
            \hline
      \end{tabular}
\end{table}

Fig. \ref{cost} gives the NMSE performance of recent deep learning methods under COST 2100 channel model. $\gamma'=\log_2(\gamma)$ is set as the x-axis for convenience. The proposed RecCsiNet shows noticeable advantages under various compression ratios. CsiNet-LSTM shares similar performance gains as RecCsiNet at some compression ratios. However, the parameter size of CsiNet-LSTM is almost five times that of ours as compared in Table \ref{TABLE: Parameters}. The parameter size of the proposed PR-RecCsiNet reduces sharply, while the performance beats CsiNet. The proposed PR-RecCsiNet can be considered as a promising option in a memory and computation limited application. Table \ref{TABLE:nmse2} gives the comparison of quantized versions under mmWave channels, which shows that our networks still outperform the conventional methods.
%Our proposed RecCsiNet and PR-RecCsiNet show pronounced advantages over various compression ratio settings. From Table \ref{TABLE: Parameters}, we can find that the parameter size of PR-RecCsiNet is on the same order of magnitude with CsiNet in \cite{CsiNet}.
%It is worth mentioning that the performance gain mainly benefits from the slowly time-varying nature of the channel.

%Because the proposed  PR-RecCsiNet compresses the channel features to a smaller dimension by a linear FCN, it could cause some information loss of the temporal features before the features are fed into LSTM. The proposed PR-RecCsiNet still outperforms CsiNet and makes sense especially in a computation-limited application.

Furthermore, we also compare the time complexity of these methods. Since there exists no GPU solution for the codebook based method, we test it on an i7-6800k CPU, and all other network based methods are tested on a Nvidia GTX 1080Ti GPU. The average processing time of AoD codebook based method is 320 ms under 8 bits/path, while CsiNet, RecCsiNet, PR-RecCsiNet and CsiNet-LSTM require 0.1 ms, 0.9 ms, 0.6 ms and 1.1 ms, respectively, when the compression ratio is 1/64. The time complexity of all the networks is almost the same, which is far less than the AoD codebook based method.

\section{Conclusion}\label{sec:conclusion}

We have proposed a new network architecture and proposed recurrent compression/uncompression modules to exploit the time-varying features in an effective manner. Apart from this, we have also provided a simple method to reduce the parameter size.
%Both the methods have the ability to outperform CS-CsiNet and CsiNet under the slowly time-varying channel.
In the letter, we limit ourselves to the design of the compression and uncompression modules. By further optimizing the feature extraction module and the channel recovery module, we expect that the encoder-decoder NN will achieve a better performance. Moreover, a direct extension to multi-user MIMO mandates assigning a Decoder Network for each user at the BS separately. Ongoing effort focuses on a more compact and integral design needs further investigation.

\end{document}